\documentclass[aps, prd, floatfix, nofootinbib, superscriptaddress, twocolumn]{revtex4-1}

\usepackage{latexsym}
\usepackage{amsmath}
\usepackage{amssymb}
\usepackage{amsfonts}
\usepackage{textcomp}
\usepackage{color}
\usepackage{CJKutf8}

\usepackage[mathscr,scaled=1.15]{urwchancal}
\DeclareFontFamily{OT1}{pzc}{}
\DeclareFontShape{OT1}{pzc}{m}{it}%
{<-> s * [1.15] pzcmi7t}{}
\DeclareMathAlphabet{\mathpzc}{OT1}{pzc}{m}{it}

\usepackage{color}

\usepackage{supertabular}
\usepackage{placeins}
\usepackage{epsfig}
\usepackage{graphicx}

\definecolor{purple}{rgb}{0.5,0,0.5}
\definecolor{blue}{rgb}{0.0,0,0.9}
\definecolor{prdblue}{rgb}{0.133,0.118,0.498}
\usepackage[colorlinks=true, pdfstartview=FitV, linkcolor=prdblue, citecolor= prdblue, urlcolor=prdblue]{hyperref}

\makeatletter

\setbox0\hbox{$\xdef\scriptratio{\strip@pt\dimexpr
    \numexpr(\sf@size*65536)/\f@size sp}$}

\newcommand{\scriptveryshortarrow}[1][3pt]{{%
    \hbox{\rule[\scriptratio\dimexpr\fontdimen22\textfont2-.2pt\relax]
               {\scriptratio\dimexpr#1\relax}{\scriptratio\dimexpr.4pt\relax}}%
   \mkern-4mu\hbox{\let\f@size\sf@size\usefont{U}{lasy}{m}{n}\symbol{41}}}}

\makeatother

\begin{document}

\begin{CJK}{UTF8}{song}

\title{$\,$\\[-6ex]\hspace*{\fill}{\normalsize{\sf\emph{Preprint no}.\ NJU-INP 050/21}}\\[1ex]
Pauli radius of the proton}

\date{2021 September 17}

\author{Zhu-Fang Cui 
}
\affiliation{School of Physics, Nanjing University, Nanjing, Jiangsu 210093, China}
\affiliation{Institute for Nonperturbative Physics, Nanjing University, Nanjing, Jiangsu 210093, China}
\author{Daniele Binosi}
\email{binosi@ectstar.eu}
\affiliation{European Centre for Theoretical Studies in Nuclear Physics
and Related Areas, Villa Tambosi, Strada delle Tabarelle 286, I-38123 Villazzano (TN), Italy}
\author{Craig D.~Roberts}
\email[]{cdroberts@nju.edu.cn}
\affiliation{School of Physics, Nanjing University, Nanjing, Jiangsu 210093, China}
\affiliation{Institute for Nonperturbative Physics, Nanjing University, Nanjing, Jiangsu 210093, China}
\author{Sebastian M.~Schmidt}
\affiliation{Helmholtz-Zentrum Dresden-Rossendorf, Dresden D-01314, Germany}
\affiliation{RWTH Aachen University, III. Physikalisches Institut B, Aachen D-52074, Germany}

\begin{abstract}
Using a procedure based on interpolation via continued fractions supplemented by statistical sampling, we analyse proton magnetic form factor data obtained via electron+proton scattering on $Q^2 \in [0.027,0.55]\,$GeV$^2$ with the goal of determining the proton magnetic radius.  The approach avoids assumptions about the function form used for data interpolation and ensuing extrapolation onto $Q^2\simeq 0$ for extraction of the form factor slope.  In this way, we find $r_M = 0.817(27)\,$fm.  Regarding the difference between proton electric and magnetic radii calculated in this way, extant data are seen to be compatible with the possibility that the slopes of the proton Dirac and Pauli form factors, $F_{1,2}(Q^2)$, are not truly independent observables; to wit, the difference $F_1^\prime(0)-F_2^\prime(0)/\kappa_p = [1+\kappa_p]/[4 m_p^2]$, \emph{viz}.\ the proton Foldy term.
\end{abstract}

\maketitle

\end{CJK}


\noindent\emph{1.$\;$Introduction} ---
The Universe is 14-billion years old.  In all that time, proton decay has not been observed.  This remarkable fact makes the proton Nature's most fundamental bound-state.  It is also a fermion; hence, owing to the spin-statistics theorem \cite{SW80}, the proton is an essentially quantum field theoretical composite object because \cite{Pauli:1940zz}: ``\ldots the connection between spin and statistics is one of the most important applications of the special relativity theory \ldots''; and quantum field theory is the only known means of unifying quantum mechanics with special relativity \cite{Weinberg:1995mt}.  The remaining Poincar\'e-invariant observable associated with an asymptotic proton state is its mass \cite{Coester:1992cg}, $m_p$, which is known with a relative uncertainty of $\sim 10^{-10}$ \cite{Zyla:2020zbs}.

The proton has electric charge $e_p=+1$, so electron+proton ($ep$) scattering has long been used to probe its dynamical properties; and being a relativistic composite object, the proton's electromagnetic current involves two form factors, Dirac -- $F_1$ and Pauli -- $F_2$ \cite{Hofstadter:1956qs, Punjabi:2015bba, Cloet:2013jya, Eichmann:2016yit, Brodsky:2020vco, Barabanov:2020jvn}:
\begin{subequations}
\begin{align}
J_\mu(Q) & = i e_p \bar u(P^\prime) \Lambda_\mu(Q) u(P)\,, \\
\Lambda_\mu(Q)& = \gamma_\mu F_1(Q^2) + \frac{1}{2 m_p} \sigma_{\mu\nu} Q_\nu F_2(Q^2)\,,
\end{align}
\end{subequations}
where $Q=P^\prime - P$, with $P$, $P^\prime$ being the incoming/outgoing proton momenta in the scattering process.  The proton electric and magnetic form factors are
\begin{equation}
G_E = F_1 - \frac{Q^2}{4 m_p^2} F_2\,, \;
G_M = F_1 + F_2\,.
\end{equation}
Three-dimensional Fourier transforms of $G_{E,M}$ were long interpreted as clean measures of the spatial distributions of electric charge and magnetisation inside the proton \cite{Sachs:1962zzc}, with the associated radii given by
\begin{equation}
\label{EqRadius}
r_{E,M}^2 = \left[\frac{-6}{G_{E,M}(Q^2)}  \frac{d}{dQ^2} G_{E,M}(Q^2)\right]_{Q^2=0}.
\end{equation}
Today, although their interpretation has changed \cite{Miller:2010nz}, the importance of measuring these key dynamical characteristics of the proton has not.

For a structureless, noninteracting fermion, the Dirac equation yields $F_2\equiv 0 \Rightarrow G_E\equiv G_M$; and perturbative corrections in quantum electrodynamics (QED) only introduce small modifications for elementary fermions \cite{Aoyama:2012wj, Aoyama:2012wk}.  Any material differences between $G_E$ and $G_M$ are marks of compositeness, and the first signal of this for the proton was found in the discovery that its anomalous magnetic moment $\kappa_p :=\mu_p-1:= F_2(0) \approx 2$ \cite{FrischStern1}.  Notwithstanding that, until the beginning of the current millennium, available $ep$ scattering data were consistent with $\mu_p G_E (Q^2)/G_M(Q^2) = 1$; and it was only the operation of an electron accelerator with the combination of high energy, luminosity, and beam polarisation that revealed $\mu_p G_E(Q^2)/G_M(Q^2) \neq 1$ \cite{Jones:1999rz}.
This being the case, then the proton electric and magnetic radii must be different; unless some dynamical mechanism leads to
\begin{equation}
\label{EqFoldy}
\frac{F_2^\prime (0)}{\kappa_p} = F_1^\prime(0) - \frac{\mu_p}{4 m_p^2} \,.
\end{equation}

Eq.\,\eqref{EqFoldy} states that the proton's Pauli and Dirac radii are not independent observables.  Rather, their difference is always positive, being fixed by the proton Foldy term \cite{Foldy:1958zz}.
%
Should this be true, then the proton's light-front transverse mean-square magnetic radius \cite{Miller:2010nz}, $b_M^\perp$, is a known quantity once its mean-square charge radius, $b_E^\perp$, is determined, with no additional input necessary; and $b_M^\perp > b_E^\perp$.
This last feature was discussed in Ref.\,\cite{Miller:2007kt}.
Of course, one could instead describe $b_E^\perp$ as the contingent quantity.

No symmetry considerations require Eq.\,\eqref{EqFoldy}; thus, the result can only emerge as a consequence of Standard Model dynamics.  One might therefore ask for the status of Eq.\,\eqref{EqFoldy} in theory; but the precision of today's theory is insufficient to deliver an answer.

\begin{figure}[t!]
	\includegraphics[width=0.999\linewidth]{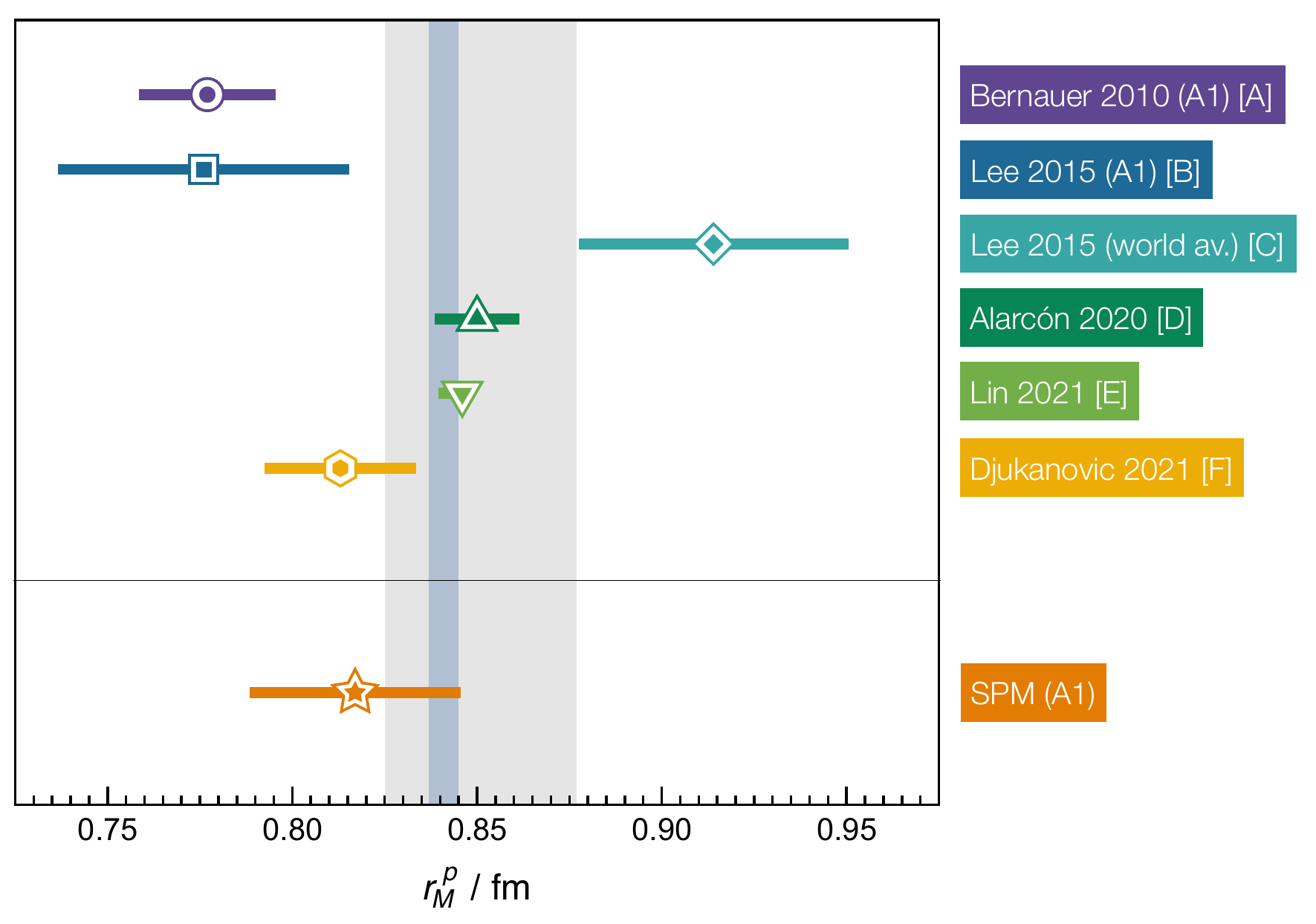}
	\caption{\label{FigMagRad}
{\it Upper panel}.  Proton $r_M$ extractions, various techniques:
[A] = Ref.\,\cite{Bernauer:2010wm, A1:2013fsc} -- A1 Collaboration;
[B] = Ref.\,\cite{Lee:2015jqa} -- A1 data;
[C] = Ref.\,\cite{Lee:2015jqa} -- world average omitting A1 data;
[D] = Ref.\,\cite{Alarcon:2020kcz} -- dispersion theory;
[E] = Ref.\,\cite{Lin:2021umz} -- dispersion theory;
[F] = Ref.\,\cite{Djukanovic:2021cgp} -- lattice quantum chromodynamics.
The light-grey band is Eq.\,\eqref{BCaverage}; within mutual uncertainties, this value agrees with $r_E$ in Eq.\,\eqref{protonradius}, which is indicated by the thinner vertical blue band.
{\it Lower panel}. Result in Eq.\,\eqref{SPMA1result}, obtained from the data in Ref.\cite{Bernauer:2010wm} using the Schlessinger Point Method (SPM) \cite{PhysRev.167.1411, Schlessinger:1966zz, Tripolt:2016cya} as described herein.}
\end{figure}

Following a decade of controversy, experiment is settling on a value for the proton electric radius \cite{Pohl:2010zza, Antognini:1900ns, Beyer:2017gug, Xiong:2019umf, Bezginov1007, Grinin1061, Pohl1:2016xoo}:
\begin{equation}
\label{protonradius}
r_E = 0.8409(4) \,{\rm fm}\,.
\end{equation}
However, as emphasised by Fig.\,\ref{FigMagRad}, the value of the proton magnetic radius is far less certain.  The most precise available data were collected by the A1 Collaboration \cite{Bernauer:2010wm, A1:2013fsc}; but as highlighted by entries [B, C] in Fig.\,\ref{FigMagRad}, applying the same analysis method to the world's data when including/excluding the A1 set can lead to mutually inconsistent results.  The particle data group (PDG) \cite{Zyla:2020zbs} quotes an average of [B, C] in Fig.\,\ref{FigMagRad}:
\begin{equation}
\label{BCaverage}
r_M = 0.851 \pm 0.026\,{\rm fm}.
\end{equation}

The proton radius controversy \cite{Gao:2021sml} demonstrated that a sound extraction of hadron radii from form factors measured in lepton+hadron scattering requires precise data, densely packed at very-low-$Q^2$.  Furthermore, in the analysis of that data, great care should be taken to eliminate all systematic error intrinsic to the choice of data fitting-function \cite{Kraus:2014qua, Lorenz:2014vha, Griffioen:2015hta, Higinbotham:2015rja, Hayward:2018qij, Zhou:2018bon, Alarcon:2018zbz, Higinbotham:2019jzd, Hammer:2019uab}.

Recently, a scheme was developed that circumvents any need for a specific choice of fitting function in analysing form factor data \cite{Cui:2021vgm, Cui:2021aee}.  Introduced in Refs.\,\cite{PhysRev.167.1411, Schlessinger:1966zz} and now often described as the statistical Schlessinger point method (SPM), it has been honed in hadron physics applications, especially those which require interpolation and reliable extrapolation, \emph{e.g}., Refs.\,\cite{Chen:2018nsg, Binosi:2018rht, Binosi:2019ecz, Eichmann:2019dts, Yao:2020vef, Yao:2021pyf, Cui:2021gzg}.  The SPM builds form-unbiased interpolations of data as the foundation for well-constrained extrapolations.  Herein, we employ it to extract $r_M$ from the A1 data \cite{Bernauer:2010wm, A1:2013fsc}.

\medskip

\noindent\emph{2.$\;$SPM and proton form factor data} ---
The data obtained by the A1 Collaboration relate simultaneously to the proton electric and magnetic radii.  They comprise $1\,400$ cross-sections measured with beam energies (in GeV): $0.18$, $0.315$, $0.45$, $0.585$, $0.72$, $0.855$.  Focusing on the electric form factor, Ref.\,\cite{Cui:2021vgm} worked with $N=40$ cross-section points in the low-$Q^2$ interval: $3.8 \times 10^{-3} \leq Q^2/{\rm GeV}^2 \leq 1.4\times10^{-2}$; and applied to this data, the SPM yielded:
\begin{equation}
\label{A1rE}
r_E^{\rm A1 - SPM} = 0.856 \pm 0.014_{\rm stat}\,{\rm fm}.
\end{equation}
Since details are available in Ref.\,\cite{Cui:2021vgm}, hereafter we only sketch the SPM via its use in extracting $r_M$.


Regarding A1 data on the magnetic form factor, one finds in  Ref.\,\cite[Supp.\,Mat.]{A1:2013fsc} $77$ points describing a Rosen\-bluth separation \cite{Rosenbluth:1950yq} of the $ep$ cross-sections into results for $G_{E,M}$ on the domain $1.5\times 10^{-2} \leq Q^2/{\rm GeV}^2 \leq 0.55$.
The first nine $G_M$  points are too noisy to be useful, only adding noise to the SPM result without altering the central value; so herein we report an analysis of the $N=68$ $G_M$ data on the domain ${\cal D}_{A1} = \{Q^2\,|\,2.7 \times 10^{-2} \leq Q^2/{\rm GeV}^2 \leq 0.55$\}.

In implementing the SPM, one constructs a prodigious number of continued fraction interpolations, every one encapsulating both local and global aspects of the curve underlying the data.  The global quality is vital because it
justifies use of the interpolations outside the domain of available data and thus enables and legitimises evaluation of the curves' slope at the origin, \emph{i.e}., use of Eq.\,\eqref{EqRadius}.

Of course, good data are statistically scattered around that curve which properly represents the form factor whose measurement is the aim; hence, they should not be directly interpolated.  We address this issue by \emph{smoothing} with a \emph{roughness penalty}, following the procedure explained in Ref.\,\cite{Reinsch:1967aa} and made concrete in Ref.\,\cite[Sec.\,3]{Cui:2021vgm}.  The approach is characterised by a well-defined optimal roughness penalty, $\epsilon$, whose value is a self-consistent output of the smoothing procedure: $\epsilon =0$ means the data are untouched by smoothing whilst $\epsilon =1$ returns a linear least-squares realignment of the data.  In the case of $r_E$, considered in Ref.\,\cite{Cui:2021vgm}, and $r_M$, herein, $\epsilon \simeq 0$.

The ${\cal D}_{A1}$ set has $N=68$ points.
From the data on ${\cal D}_{A1}$, we randomly select $7 \leq L \leq 19$ elements.
With a given value of $L$, one can generate $C(N,L)$ different interpolating functions, \emph{i.e}., O$(9.7 \times 10^8 -3.3 \times 10^{15})$ interpolators.
From the collection obtained using a given value of $L$, we select the first $5\,000$ that are smooth on the entire $Q^2$ domain covered by the data.
Every interpolating function defines an extrapolation to $Q^2=0$, wherefrom $r_M$ can be obtained via Eq.\,\eqref{EqRadius}.  So, for each value of $L$, $r_M$ is determined by the average of all results computed from the $5\,000$ curves.

To estimate the uncertainty in a particular SPM-determined value of $r_M$, one must first reckon with the experimental error in the data.  This is achieved herein by using a statistical bootstrap procedure \cite{10.5555/1403886}.  We generate $1\,000$ copies of the set by replacing each element by a new point, distributed randomly about the mean defined by the element itself with variance equal to its associated error.
Additionally, since $L$ is changed, there is a second error source $\sigma_{{\delta\!L}}$, which can be estimated by shifting $L\to L'$, repeating all steps for this new $L'$-value, and  computing the standard deviation of the distribution of $r^{L}_M$ for different $L$ values.

Accordingly, one arrives at the SPM result: $r_M\pm\sigma_r$;
\begin{align}
	&r_M=\sum_{j=1}^{4}\frac{r^{{L_j}}_M}{4};&
&\sigma_r=\Bigg[\sum_{j=1}^{4} \frac{(\sigma^{{L_j}}_r)^2}{4^2}
+\sigma_{{\delta\!L}}^2\Bigg]^{\frac12}.
	\label{SPMrp}
\end{align}
For reasons explained below, we calculate results for each one of the values $L$ in ${\cal S}_L=\{L_j=3+4j\,\vert\ j = 1,2,3,4\}$.  Thus, we have $20$--million results for $r_M$, each computed from an independent interpolation. For all $j$ values in the range specified above, $\sigma_{{\delta\!L}}\ll \sigma_{r}^{ L_j}$; hence, the result is largely independent of the chosen value of $L$.

\medskip

\noindent\emph{3.$\;$Validation of SPM extrapolation on A1 data} ---
The keystone of any SPM data analysis is validation on the set.  Herein, this means demonstrating that the radius returned by the SPM is reliable when used in connection with data on ${\cal D}_{A1}$.  We accomplish this using the approach explained in Ref.\,\cite[Supp.\,Mat.]{Cui:2021vgm}.

\hspace*{-0.5\parindent}{\sf Step 1}.
The A1 Collaboration employed ten distinct data fitting models as the means by which to infer a result for $r_M$ \cite{A1:2013fsc}.  From this collection, we selected the seven that are most straightforward to implement: (\emph{i}) dipole; (\emph{ii}) double dipole; (\emph{iii}) polynomial; (\emph{iv}) polynomial plus dipole; (\emph{v}) polynomial times dipole; (\emph{vi}) inverse polynomial; and (\emph{vii}) Friedrich-Walcher parametrisation \cite{Friedrich:2003iz}.  Each of these models is connected with a distinct magnetic radius \cite{A1:2013fsc}, which defines the target input radius, $r_M^{\rm input}$, that the SPM should recover (in fm):
$r_M^{i} = 0.85\,$;
$r_M^{ii} = 0.838761\,$;
$r_M^{iii} = 0.779342\,$;
$r_M^{iv} = 0.778892\,$;
$r_M^{v} = 0.775504\,$;
$r_M^{vi} = 0.768562\,$;
$r_M^{vii} = 0.807416\,$.
Using every one of these models, we generated replicas of the A1 data built from the model values of $G_M$ evaluated at the $Q^2$ points in ${\cal D}_{A1}$.  The character of real data was modelled by scattering the values in response to fluctuations drawn according to a normal distribution.

\hspace*{-0.5\parindent}{\sf Step 2}.
Treating each of the seven sets as real, we determined the radius as described above.
(\emph{I}) Generate $10^3$ replicas.
(\emph{II}) Smooth each replica with the optimal parameter.
(\emph{III}) Use the SPM to obtain $r^{L_j}_M$ and $\sigma^{L_j}_M$, varying the number of input points $\{L_j=3+4j\,\vert\ j = 1,2,3,4\}$.
(\emph{IV}) Compute the final SPM result via Eq.\,\eqref{SPMrp} and compare it with the input value.

The results of validation are summarised thus.
\begin{description}
\item[A]
For a given value of $L\in {\cal S}_L$ and all seven forms of fitting model, the distribution of SPM-extracted radii is a Gaussian, which is centred on the radius input value and whose characteristics are practically independent of $L$: in these cases, $\sigma_{{\delta\!L}}\ll \sigma^{{L_j}}_r$.
We also considered $L \in {\cal O}_L=\{ M_{ji} = 3 + 4 j + i\,\vert\, j=1,2,3,i=1,2,3\}$.
In these instances, however, one does not recover a satisfactory Gaussian distribution.  So, we omit them from further consideration; to wit, we allow the validation process to impose a subset of $L$-values on our analysis, \emph{viz}.\ $L\in {\cal S}_L$.

\item[B]
Defining the bias: \mbox{$\delta r_M=r_M-r^{\rm input}_M$}, then the SPM extraction of the proton magnetic radius is robust: typically, $|\delta r_M| \lesssim \sigma_r$, where $\sigma_r$ is the standard error in Eq.\,\eqref{SPMrp}.  The inverse polynomial model is the worst case.
These remarks are illustrated in Fig.\,\ref{FValidate}, which depicts $\delta r_M$ as obtained using the SPM to recover $r_M$ from each one of the seven $G_M$ generators with their distinct $r_M^{\rm input}$ values.

\begin{figure}[t!]
	\includegraphics[width=0.999\linewidth]{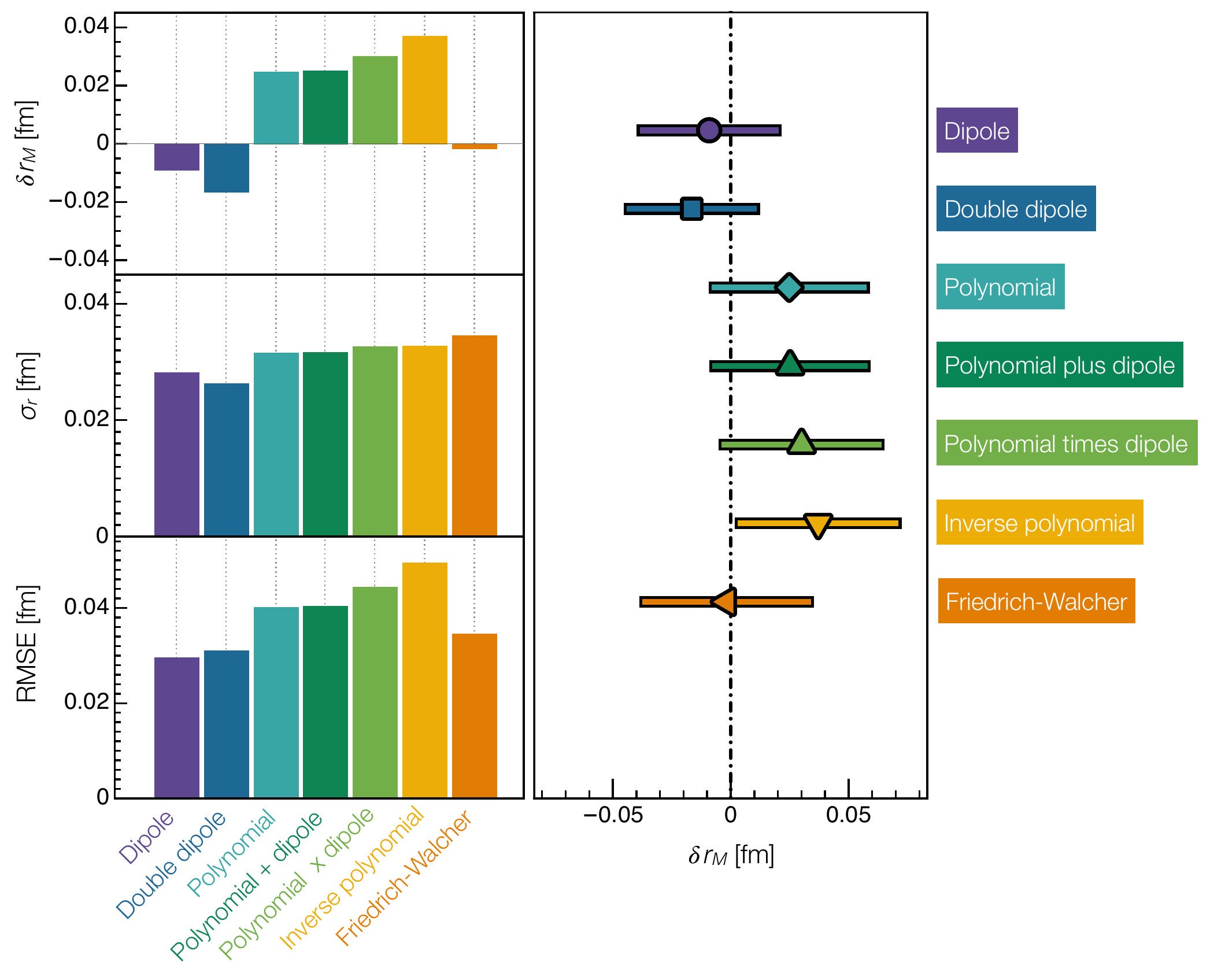}
\caption{\label{FValidate}
\emph{Top and right panels} -- bias, $\delta r_M=r_M-r_M^{\rm input}$, where the values of $r_M^{\rm input}$ are listed in {\sf Step 1}; \emph{middle panel} -- standard error, Eq.\,\eqref{SPMrp}; and \emph{bottom panel} -- RMSE, Eq.\,\eqref{EqRMSE}, for SPM extractions of the proton magnetic radius from $5\,000$ replicas built using the seven models employed by the A1 Collaboration \cite{A1:2013fsc}.
}
\end{figure}

\item[C]
Considering the root mean square error:
\begin{equation}
	\mathrm{RMSE}=\sqrt{(\delta r_M)^2+\sigma_r^2}\,,
\label{EqRMSE}
\end{equation}
then, as revealed in Fig.\,\ref{FValidate}, the SPM produces RMSE values which are approximatively independent of the model used to build the replicas.  It therefore satisfies a standard ``goodness of fit'' criterion \cite{Yan:2018bez}.  Hence, the SPM extraction of $r_M$ may objectively be judged to return a reliable expression of the information contained in the data.

\end{description}

\medskip

\noindent\emph{4.$\;$Proton magnetic radius from A1 data} ---
With the SPM now shown to return a reliable radius result from A1-like data, with its quoted errors, we turn to the real data on ${\cal D}_{A1}$ and obtain:
\begin{equation}
\label{SPMA1result}
r_M^{\rm A1 - SPM} = 0.817 \pm 0.027_{\rm stat}\,.
\end{equation}
As evident in Fig.\,\ref{FigMagRad}, within mutual uncertainties, this result agrees with that reported in Ref.\,\cite{Bernauer:2010wm}, but the central value is 5.1\% larger.  This contrasts with our kindred extraction of $r_E$ from A1 data, discussed elsewhere \cite{Cui:2021vgm}.  In that case, the central result from Eq.\,\eqref{A1rE} is 2.6\% lower.

\begin{figure}[t!]

	\includegraphics[width=0.8\linewidth]{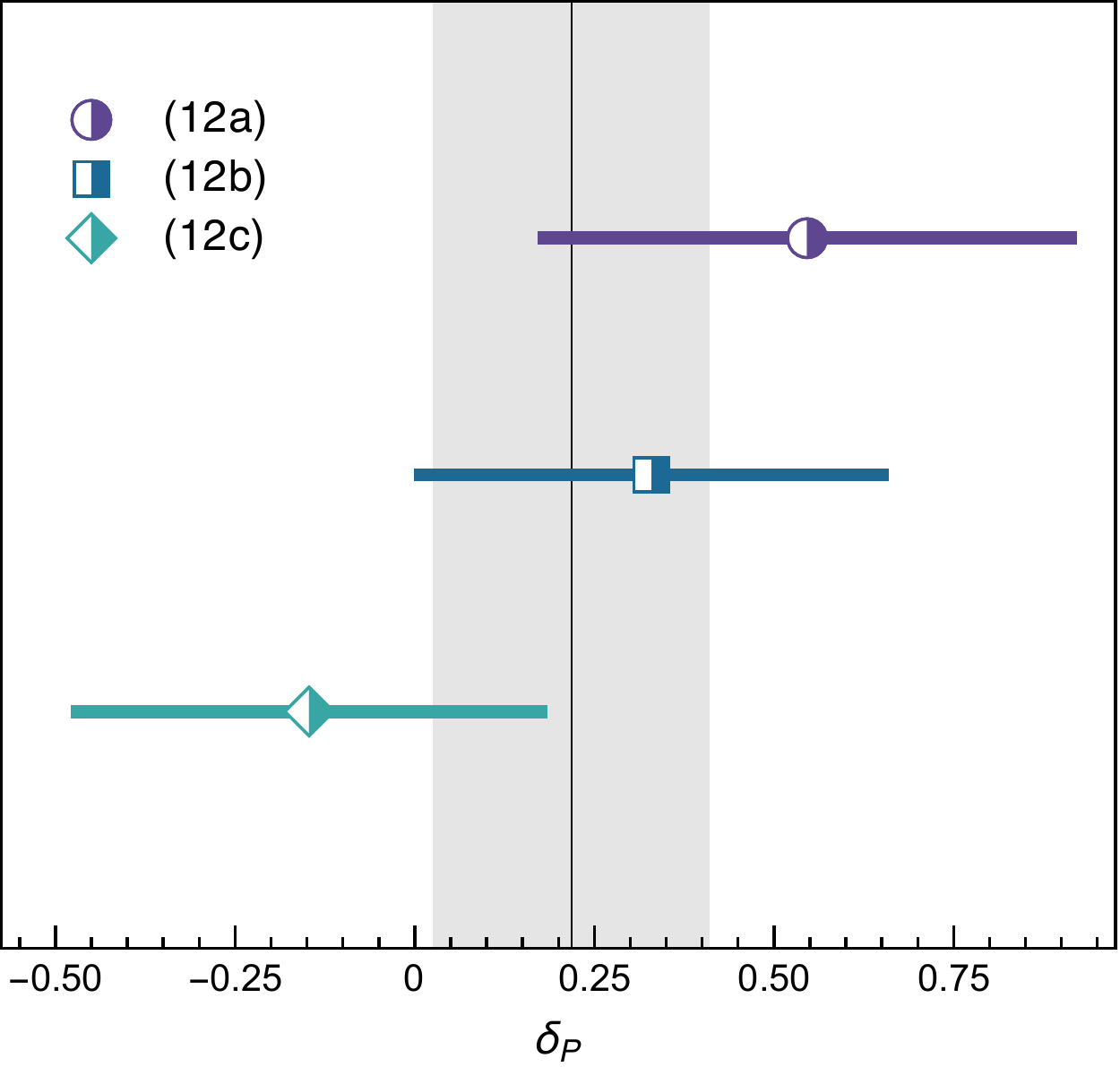}
\caption{\label{FdeltaP}
Experiment-based results for deviations ($\delta_P \neq 0$) from Eq.\,\eqref{EqFoldy} as listed in Eqs.\,\eqref{EqDelta1}.  The vertical grey line and associated band is the uncertainty-weighted average.
}
\end{figure}

These differences have meaning in connection with Eq.\,\eqref{EqFoldy}.  Writing
\begin{equation}
\label{EqFoldyApprox}
F_1^\prime(0) = {\mathpzc d}_1  \frac{\mu_p}{4 m_p^2}\,,  \;
\frac{1}{\kappa_p}F_2^\prime(0)  = [{\mathpzc d}_1-1-\delta_P]  \frac{\mu_p}{4 m_p^2}\,,
\end{equation}
then $b_M^\perp>b_E^\perp \; \forall \delta_P>-1$ and Eq.\,\eqref{EqFoldy} is recovered with $\delta_P=0$.
Working with the radii discussed above, the combinations
Eqs.\,(\ref{A1rE},\ref{SPMA1result}),
Eqs.\,(\ref{protonradius},\ref{SPMA1result}),
Eqs.\,(\ref{protonradius},\ref{BCaverage}) yield the results drawn in Fig.\,\ref{FdeltaP}; namely:
\begin{subequations}
\label{EqDelta1}
\begin{align}
\delta_P^{{\rm A1}_{\rm SPM}} &= \phantom{-}0.546 \pm 0.366\,, \\
\delta_P^{\rm {\rm A1}_{\rm SPM} + PDG} &= \phantom{-}0.329 \pm 0.321 \,,\\
\delta_P^{\rm PDG } &= -0.147 \pm 0.322 \,,
\end{align}
\end{subequations}
the uncertainty-weighted average of which is $\delta_P=0.218 \pm 0.193$.

Since the low-$Q^2$ behaviour of $[1-\mu_p G_E(Q^2)/G_M(Q^2)]$ is proportional to $\delta_P$, then greater accuracy might be achieved through new high-precision low-$Q^2$ polarisation-transfer measurements of $\mu_p G_E(Q^2)/G_M(Q^2)$, as also remarked elsewhere \cite{Miller:2007kt}.

\medskip

\noindent\emph{5.$\;$Discussion and Conclusions} ---
Fig.\,\ref{FdeltaP} reveals that analyses of available data are consistent with Eq.\,\eqref{EqFoldy} at a 26\% confidence level.  Namely, they suggest with this probability that the proton's Dirac and Pauli radii are not truly independent observables; but, instead, their difference is completely fixed by the proton Foldy term  \cite{Foldy:1958zz}. At worst, Eq.\,\eqref{EqFoldy} is established as a useful approximation.  Moreover, the analyses indicate with great certainty that the proton's Pauli radius is larger than its Dirac radius.   These conclusions will serve as valuable constraints on pictures of proton structure.

Their importance is elevated by the fact that any discussion of the ``size'' of the ``absolutely'' stable proton leads naturally to a discussion of confinement \cite{Krein:1990sf, Jaffe:Clay}.  Namely, the empirical fact that the gluon and quark quanta associated with the fields used to define quantum chromodynamics (QCD), the strong interaction part of the Standard Model, have never been observed in a detector.  Confinement has been linked with the phenomenon of emergent hadron mass (EHM) \cite{Aguilar:2019teb, Roberts:2020hiw, Roberts:2021xnz}, \emph{viz}.\ the appearance of nuclear-size masses for strong interaction systems comprised of massless field quanta.

The QCD proton is just such a bound state: seeded by three light valence quarks -- two $u$ quarks and one $d$ quark, which are all massless in the absence of Higgs boson couplings into QCD.  Symmetries in quantum field theory entail that a massless fermion cannot possess an anomalous magnetic moment \cite{Singh:1985sg, Bicudo:1998qb, Chang:2010hb}; and, in the absence of any strong corrections, the anomalous magnetic moment of a light fermion is small.  Yet, the proton's Pauli form factor, which expresses all its anomalous magnetic properties, is large on $Q^2\simeq 0$ and has a radius fixed (perhaps approximately) by Eq.\,\eqref{EqFoldy}.

An explanation for the infrared scale of $F_2$-proton may be found in EHM \cite{Aguilar:2019teb, Roberts:2020hiw, Roberts:2021xnz}: realised most fundamentally in QCD via the dynamical generation of running masses for gluons -- QCD's gauge bosons -- and quarks \cite{Aguilar:2015bud, Fischer:2018sdj, Roberts:2021nhw}; and expressed in manifold corollaries \cite{Qin:2020jig, Roberts:2021nhw, Arrington:2021biu}.  For example, the running masses that owe to EHM are large at infrared momenta, with that of the quark being $\sim m_p/3$, and the same phenomenon also endows dressed light quarks with anomalous magnetic moments that are $\sim \kappa_p/3$ in the infrared.  This explains the size of $F_2(0)$.  Given then Eq.\,\eqref{EqFoldy}, even as an approximation, the amount by which the proton's light-front transverse mean-square magnetic radius exceeds the analogous charge radius is also seen to be (predominantly) determined by EHM.
(Numerous theory analyses indicate that EHM also strongly influences $F_1(Q^2)$, with effects flowing into both the proton wave function and electromagnetic current via the running light quark masses \cite{Cloet:2013jya, Eichmann:2016yit, Brodsky:2020vco, Barabanov:2020jvn}.)

This discussion suggests that the scales underlying the proton's anomalous magnetisation form factor may be attributed to EHM.  Lacking, however, is quantitatively precise theory that can match QCD with Eqs.\,\eqref{EqFoldy}, \eqref{EqDelta1}.  There are many possible contributions to $F_2(Q^2\simeq 0)$, including \cite{Punjabi:2015bba, Cloet:2013jya, Eichmann:2016yit, Brodsky:2020vco, Barabanov:2020jvn, Giannini:2015zia, Sufian:2016hwn, Xu:2019ilh, Mondal:2019jdg, Cui:2020rmu, Xu:2021mju}: quark anomalous magnetic moments, quark orbital angular momentum within the proton, meson cloud contributions to one or both, etc.; and related influences on $F_1(Q^2\simeq 0)$. An insightful understanding of the accuracy of Eq.\,\eqref{EqFoldy} will demand an explanation of how all these factors might combine to produce a simple algebraic outcome; and, in step with that, experiments and associated analyses that deliver a precise result for $r_M$ so that a veracious benchmark is provided.

%
\medskip
\noindent\emph{Acknowledgments}.
We are thankful for constructive comments from L.~Chang, J.~Friedrich, D.\,W.~Higinbotham, R.\,J.~Holt, V.~Mokeev, W.-D.~Nowak and J.~Segovia.
%
Use of the computer clusters at the Nanjing University Institute for Nonperturbative Physics is gratefully acknowledged.
Work supported by:
National Natural Science Foundation of China (grant nos.\,12135007 and 11805097);
Jiangsu Provincial Natural Science Foundation of China (grant no.\,BK20180323);
and STRONG-2020 ``The strong interaction at the frontier of knowledge: fundamental research and applications'' which received funding from the European Union's Horizon 2020 research and innovation programme (grant no.\,824093).


\end{document}